\documentclass[twocolumn]{revtex4-1}
\usepackage{graphicx}
\usepackage{dcolumn}
\usepackage{amsmath}
\usepackage{color}

\begin{document}

\title{Massless and quantized modes of kinks
in the phase space of superconducting gaps}

\author{Takashi Yanagisawa, Izumi Hase and Yasumoto Tanaka}

\affiliation{Electronics and Photonics Research Institute, National Institute of
Advanced Industrial Science and Technology, 1-1-1 Umezono, Tsukuba,
Ibaraki 305-8568, Japan}

%\date{}

\begin{abstract}
We investigated quantized modes of kinks in the phase space of
superconducting gaps in a superconductor with multiple gaps.
The kink is described by the sine-Gordon model in a two-gap superconductor
and by the double sine-Gordon model in a three-gap superconductor.
A fractional-flux vortex exists at the edge of the kink, and 
a fractional-flux vortex will be stable in a three-gap superconductor
with time-reversal symmetry breaking.
The kink and fractional-flux vortex exhibit massless modes as a sliding motion.
We show further that there are one zero-energy mode (massless mode) and quantized excitation
modes in kinks, which are characteristic features of multi-gap superconductors.  
The equation of quantized modes for the double sine-Gordon
model is solved numerically.
The correction to the ground-state energy is calculated based on the renormalization
theory.
\end{abstract}


\maketitle

\section{Introduction}

The study of multiband superconductors started from old works by
Moskalenko\cite{mos59},
Suhl et al.\cite{suh59}, Peretti\cite{per62} and Kondo\cite{kon63}.
They were regarded as a generalization of the Bardeen-Cooper-Schrieffer
(BCS) theory\cite{bcs57} to a multiband superconductor\cite{bin80}.
They are, however, not only a simple generalization but also
contain many interesting and fruitful properties.
There appear many interesting properties in multi-component (multi-band) 
superconductors which are not found in a single-band superconductor. 
They are, for example, time-reversal symmetry 
breaking 
(TRSB)\cite{tan01,tan09,sta10,tan10a,tan10b,dia11,yan12,hu12,sta12,pla12,
mai13,wil13,gan14,yer15,tan15c},
the existence of massless modes\cite{tan11,yan13,lin12,kob13,koy14,yan14,tan15},
the existence of vortices with fractionally 
quantized-flux\cite{yan12,izy90,vol09,tan02,bab02,kup11,gar11,gar13,smi05,tan18b},
unconventional isotope effect\cite{cho09,shi09,yan09},
and a new type of superconductors called the type-1.5 
superconductivity\cite{mos09,sil11,car11}.
The existence of fractional quantum-flux vortices has been examined
theoretically and is now an interesting subject.
It is expected to be hard to observe fractional-flux vortices in 
real superconductors
and there have been several experimental attempts to demonstrate 
them\cite{blu06,cri07,gui08,lua09,tan15b}. 
There are theoretical and experimental proposals concerning
the stabilization of fractionally quantized flux
in a two-band superconductor\cite{kup11,pin12,lin13,tan17}.
Recently, the experimental observation of fractional vortices in a
thin superconducting bi-layer was reported as a magnetic flux distribution
image taken by a scanning superconducting quantum interference device
(SQUID) microscope.\cite{tan18}.  It is important to study physical
phenomena concerning fractional-flux vortices.

In a two-band superconductor, a solution of the half-flux vortex
exists.  The dynamics of phase difference $\varphi$ between two gaps
is described by the sine-Gordon model\cite{raj82,wei12,yan16,yan17,yan18}:
\begin{equation}
\nabla^2\varphi-\alpha\sin\varphi=0,
\end{equation}
where $\alpha$ is a constant proportional to the Josephson
coupling constant.
We adopt that the phase difference $\varphi$ has spatial
dependence only in one direction $x$.
We have a kink solution for the boundary condition that
$\varphi\rightarrow 0$ as $x\rightarrow -\infty$ and
$\varphi\rightarrow 2\pi$ as $x\rightarrow\infty$.
We assume that we have two equivalent bands with the same
intraband coupling constants.
A half-quantum flux vortex exists at the edge of the kink\cite{yan13}.
This is shown in Fig.1, where the phase variables $\theta_1$ and $\theta_2$
change across the kink from 0 to $\pi$ and 0 to $-\pi$, respectively.
A net-change of $\theta_1$ is $2\pi$ by a counterclockwise encirclement
of the vortex, and that of $\theta_2$ vanishes.
Here the magnetic field is perpendicular 
to the plane. 
The phase difference $\varphi$ changes from 0 to $2\pi$ as going
across the kink.
The other end of the kink is at the boundary of a superconductor.
Two half-flux vortices are connected by the kink solution of
the sine-Gordon equation.
In a superconductor in three dimensions, the kink
forms a surface (two-dimensional plane) between two fractional
vortices.

Let us discuss a stability of fractional-flux vortices.
We need energy to form the kink between fractional vortices,
where the energy is proportional to the square root of the Josephson
coupling\cite{yan12}.
Since the energy of magnetic field is proportional to the
square of the flux, the total energy of a pair of half-flux vortices
is just half of that of a vortex with the unit flux.
However, in general, the energy of the kink is larger than
that, and thus the pair will be glued to be a 
unit-flux vortex\cite{tan07}.
A pair of half-flux and anti-half-flux vortices will vanish
completely because of the magnetic energy $\propto \phi_0^2$. 
%Thus the fractional vortices are in general unstable and  
%appear as excited states at
%finite temperature.

Our discussion is based on the effective action
derived from the BCS theory for a multi-component superconductor\cite{yan17b}.  
We focus on excitation
modes concerning kinks and fractional vortices since the Nambu-Goldstone
and Leggett modes have been studied by many authors\cite{yan13,lin12,kob13,
koy14,yan14,tan15,sha02,yan17c,yan17b}.
The derived action contains the second derivative with respect to time.
In this paper we do not introduce the dissipation term which is the
first derivative term with respect to time\cite{gur06}.

In this paper, we first consider a stability of fractional-flux vortices
in multi-component superconductors.
We argue that fractional-flux vortices indeed
can be stabilized in a three-band superconductor with time-reversal
symmetry breaking. 
Second, we examine excitation modes of kinks in two- and three-gap
superconductors, and show that there are zero-energy mode and quantized
excitation modes.  Generally, there are zero-energy modes when there are
fractional vortices and kinks.
The paper is organized as follows.
In Section II, we discuss a stability of fractional-flux vortices
on a kink in a superconductor with time-reversal
symmetry breaking.
In Section III, we investigate the quantization of a kink solution
of the sine-Gordon model for two- and three-gap superconductors.
We give a summary in the last Section.

\vspace{1.0cm}
\begin{figure}[htbp]
\begin{center}
\includegraphics[height=6.0cm,angle=90]{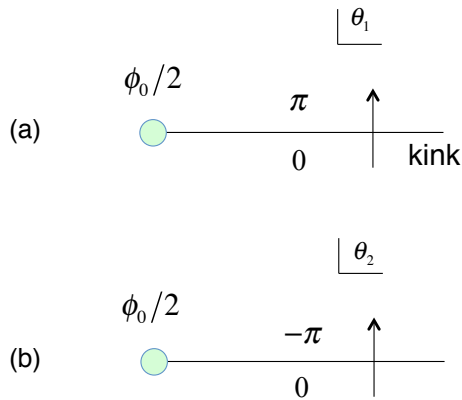}
\caption{
Half-quantum flux vortex with a kink (line singularity).
The phase variables $\theta_1$ and $\theta_2$ change from
(a) 0 to $\pi$ and (b) 0 to $-\pi$  across the kink, respectively.
}
\label{kink}
\end{center}
\end{figure}

\section{Kinks (defects) and states of fractional-flux vortices}

In a two-component (two-gap) superconductor, there may
exist a kink that runs from one end of the boundary to the other end 
as shown in Fig.2(a).
We call this type of kink the transverse kink in a superconductor.
%Let us represent the state without a fractional vortex by a set
%$\{ {\rm point} \}$, while a vortex state is by ${\bf R}$.
%The existence of the transverse kink  may be interpreted as
%\begin{equation}
%K^0(\{{\rm point}\})= {\bf Z}.
%\end{equation}
This is also the case for superconductors with bands more than
two.  We write the Josephson potential in the form
\begin{equation}
V= \sum_{i>j}\Gamma_{ij}\cos(\theta_i-\theta_j),
\end{equation}
where $\theta_j$ is the phase of the gap
function in the $j$-th gap for $j=1,2,\cdots,N$
($N$ is the number of gaps) and $\Gamma_{ij}$ are constants.
We assume that $\Gamma_{ij}=\Gamma_{ji}$.
When all the $\Gamma_{ij}$ are negative, the ground state is
at $\theta_i-\theta_j=0$ mod$2\pi$.
We have a kink that connects two states, for example,
$(\theta_1,\theta_2,\cdots)=(0,0,\cdots,0)$ and
$(\pi,-\pi,\pi,-\pi,\cdots)$.
A fractional-flux vortex can exist at the edge of the kink
as in the two-band case.

We assume that the kink surface is a plane and the magnetic
field is applied being parallel to the plane.
When the vortex with unit flux $\phi_0$ exists just at the
plane (Fig.2(b)), the flux $\phi_0$ is separated into two
half-flux vortices (Fig.2(c)).
The kink will disappear to reduce the energy of the kink. 

Let us turn to examine the time reversal symmetry broken superconducting
state.  In a three-band superconductor, the time reversal symmetry
(TRS)
is broken when $\Gamma_{12}\Gamma_{23}\Gamma_{31}>0$. 
In this case, a fractional vortex has two kinks because we
have two phase-difference modes\cite{yan12}.
When three bands are equivalent, a fractional vortex with the
flux $\phi_0/3$ exists with two kinks. 
It is possible to have a pair of fractional vortices which
is the bound state of
fractional vortices that corresponds to the meson under the
duality transformation between charge and magnetic flux\cite{yan12}.
The attractive interaction works between two vortices
because the energy of kink is proportional to the distance
between fractional vortices.

A fractional-vortex pair on the kink can be stable in a superconductor
with time reversal symmetry breaking.
This is shown in Fig.3.
Let us consider a three-band superconductor with
equivalent bands and thus the Josephson potential $V$
has $|\Gamma_{12}|=|\Gamma_{23}|=|\Gamma_{31}|$ with
$\Gamma_{12}\Gamma_{23}\Gamma_{31}>0$. 
The ground state has a $2\pi/3$ structure in this case.
We adopt that there is a kink that connects two states with
$(\theta_1,\theta_2,\theta_3)= (2\pi/3,0,-2\pi/3)$ and
$(\theta_1,\theta_2,\theta_3)= (-2\pi/3,0,2\pi/3)$.
Now we put a vortex with unit flux $\phi_0$ on the kink.
This vortex will be separated into two fractional vortices
as shown in Fig.3(c) because the flux energy proportional
to $(\phi_0/3)^2+(2\phi_0/3)^2$ is less that of the unit
flux proportional to $\phi_0^2$.
The energy of the kink remains the same after the
separation of vortices.  The phases of gaps are shown
in Fig.3(c).  Thus a pair of vortices can be stabilized
in a TRSB superconductor.
This is because the kink structure is not trivial in
the TRSB case compared to that in the TRS case.

\vspace{0.5cm}
\begin{figure}[htbp]
\begin{center}
\includegraphics[height=7.0cm]{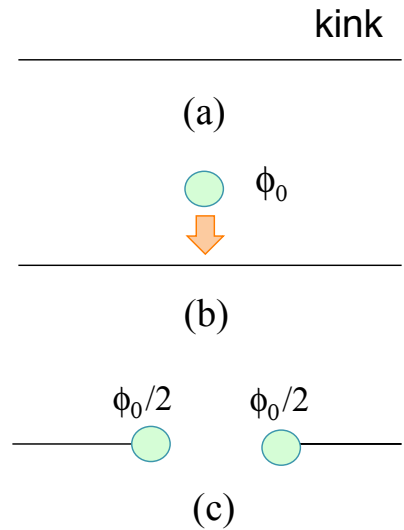}
\caption{ Kink and vortices on the plane perpendicular
to the magnetic field.
(a) A kink from one end to another end in a two-band 
superconductor.
(b) A unit flux vortex approaching the kink.
(c) The unit flux on the kink is separated into a pair of
half-flux vortices and will disappear.
}
\label{fluxpair3}
\end{center}
\end{figure}

\vspace{1.0cm}
\begin{figure}[htbp]
\begin{center}
\includegraphics[height=7.0cm]{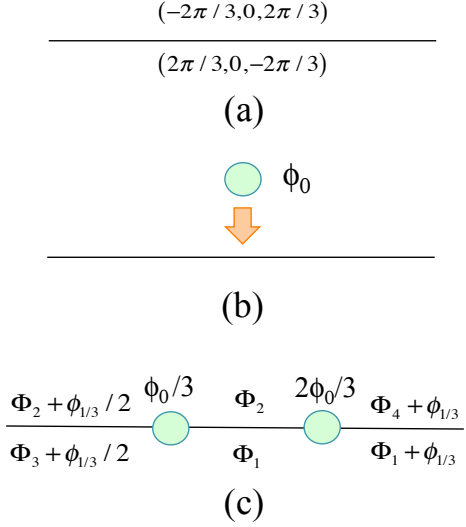}
\caption{
(a) A kink with in a three-band superconductor 
with equivalent bands.  We assume that the time reversal symmetry
is broken  The phases of gap functions below the kink is
$\Phi_1\equiv (2\pi/3,0,-2\pi/3)$ and above the kink is
$\Phi_2\equiv (-2\pi/3,0,2\pi/3)$.
Two states with $\Phi_1$ and $\Phi_2$ are degenerate with
the same energy.
(b) A vortex with the unit flux approaching the kink in (a).
(c) The unit-flux vortex in (b) is separated into two fractional
vortices on the kink.  We defined
$\Phi_3\equiv (0,-2\pi/3,2\pi/3)$,
$\Phi_4\equiv (0,2\pi/3,-2\pi/3)$ and $\phi_{1/3}=(2\pi/3,2\pi/3,2\pi/3)$.
Note that $\Phi_2=\Phi_4+2\phi_{1/3}$ mod$2\pi$.
The phases of gaps are shown in the figure.
}
\label{fluxpair5}
\end{center}
\end{figure}

%In a three-band superconductor with time-reversal symmetry breaking, 
%a pair of fractional vortices with flux $\phi_0/3$ and $2\phi_0/3$
%can be stable.  The space of states $H_1$ consists of  two states:
%\begin{equation}
%H_1 =\{ E_{1/3,2/3}, E_{2/3,1/3}\} \cong {\bf Z}_2,,
%\end{equation}
%where $E_{1/3,2/3}$ and
%$E_{2/3,1/3}$ indicate the states in Figs.4(a) and (b), respectively.
%Two states are degenerate.
%When a vortex with the unit flux is trapped in the kink, the vortex
%is decomposed into a pair of vortices with fractional flux
%$\phi_0/3$ and $2\phi_0/3$. 
%A state with two pairs of fractional-flux vortices is also possible 
%which is shown in Fig.4(c).  In this case, we have 4 states.
%In general, we have the space of states $H_n$ with $n$ pairs of fractional-flux 
%vortices.  The space $H_n$ has $2^n$ elements since
%\begin{equation}
%H_n \cong {\bf Z}_2\times \cdots \times {\bf Z}_2,
%\end{equation}
%where ${\bf Z}_2$ appears $n$ times in the right-hand side. 

%\vspace{0.5cm}
%\begin{figure}[htbp]
%\begin{center}
%\includegraphics[height=7.5cm]{flux-pair-3band.eps}
%\caption{
%Fractional-flux vortices in the kink in a three-band superconductor.
%We have two states shown in (a) and (b) when there is a pair of vortices.
%The figure (c) indicates a state when there are two pairs of fractional-flux
%vortices.  There $2^2$ states in this case.
%}
%\label{fluxpair-3band}
%\end{center}
%\end{figure}

\section{Quantization of kinks}
\subsection{Multi-component effective model}

We consider the Hamiltonian given by
\begin{eqnarray}
H&=& \sum_{n\sigma}\int d{\bf r}\psi_{n\sigma}^{\dag}({\bf r})
K_n({\bf r})\psi_{n\sigma}({\bf r}) \nonumber\\
&& -\sum_{nm}g_{nm}\int d{\bf r}\psi_{n\uparrow}^{\dag}({\bf r})
\psi_{n\downarrow}^{\dag}({\bf r})\psi_{m\downarrow}({\bf r})
\psi_{m\uparrow}({\bf r}),
\end{eqnarray}
where $n$ and $m$ stands for band indices and $K_n({\bf r})$ indicates
$K_n({\bf r})={\bf p}^2/(2m_n)-\mu$ where $\mu$ is the chemical potential
and $m_n$ is the mass of electrons in the $n$-th band.

For a two-band superconductor, the Lagrangian density concerning
phase variables is given by\cite{yan17b}
\begin{eqnarray}
\mathcal{L}&=& \frac{\rho_1}{4}(\partial_t\theta_1)^2
+ \frac{\rho_2}{4}(\partial_t\theta_2)^2
-\frac{n_1}{8m_1}(\nabla\theta_1)^2-\frac{n_2}{8m_2}(\nabla\theta_2)^2
\nonumber\\
&& - 2\gamma_{12}\Delta_1\Delta_2\cos(\theta_1-\theta_2),
\end{eqnarray}
where $\theta_i$ is the phase of the $i$-th gap $\Delta_i$, $\rho_i$ is
the density of states in the $i$-th band, $n_i$ is the electron density
and $m_i$ indicates the mass of electrons in the $i$-th band.
$\gamma_{12}$ is the Josephson coupling constant given by
$\gamma_{12}=(g^{-1})_{12}$ for the BCS coupling constants $g=(g_{ij})$.
We adopt that $\gamma_{12}<0$ and put $K_i=n_i/(2m_i)$.

We define the phase difference $\varphi$ and the total phase $\phi$ as
$\varphi = \theta_1-\theta_2$ and
$\phi = \theta_1+a\theta_2$
where $a$ is a real constant.  We put $v_0$ as
\begin{equation}
v_0^2= \frac{K_1+K_2}{\rho_1+\rho_2}.
\end{equation}
The Lagrangian is written as
\begin{eqnarray}
\mathcal{L}&=& \frac{K_1+K_2}{4(a+1)^2}\left( \frac{1}{v_0^2}
(\partial_t\phi)^2-(\nabla\phi)^2\right) \nonumber\\
&& +\frac{1}{2(a+1)^2}\left( (a\rho_1-\rho_2)\partial_t\phi\partial_t\varphi
-(aK_1-K_2)\nabla\phi\nabla\cdot\varphi\right)
\nonumber\\
&& +\frac{1}{4(a+1)^2}\left( (a^2\rho_1+\rho_2)(\partial_t\varphi)^2
-(a^2K_1+K_2)(\nabla\varphi)^2 \right) \nonumber\\
&& - 2\gamma_{12}\Delta_1\Delta_2\cos(\theta_1-\theta_2).
\end{eqnarray}
Here we set $aK_1=K_2$ and integrate out the field $\phi$ to obtain
the effective action $\mathcal{L}_{\varphi}$ for $\varphi$.
We have
\begin{eqnarray}
\mathcal{L}_{\varphi}&=& \frac{\rho_1\rho_2}{4(\rho_1+\rho_2)}(\partial_t\varphi)^2
-\frac{K_1K_2}{4(K_1+K_2)}(\nabla\varphi)^2 \nonumber\\
&& -2\gamma_{12}\Delta_1\Delta_2\cos(\varphi) \nonumber\\
&=& \frac{K_1K_2}{4(K_1+K_2)}\left( \frac{1}{v^2}(\partial_t\varphi)^2
-(\nabla\varphi)^2 \right) \nonumber\\ 
&& -2\gamma_{12}\Delta_1\Delta_2\cos(\varphi) \nonumber\\
&=& A\left( \frac{1}{2v^2}(\partial_t\varphi)^2-\frac{1}{2}(\nabla\varphi)^2
+\alpha\cos(\varphi) \right),
\end{eqnarray}
where $v$ is the velocity defined by
\begin{equation}
v^2 = \frac{\rho_1+\rho_2}{\rho_1\rho_2}\frac{K_1K_2}{K_1+K_2}
= \left( \frac{1}{\rho_1}+\frac{1}{\rho_2} \right)
\left( \frac{1}{K_1}+\frac{1}{K_2} \right)^{-1},
\end{equation}
and we put 
\begin{eqnarray}
A &=& \frac{K_1K_2}{2(K_1+K_2)},\\
\alpha &=& -\frac{K_1+K_2}{K_1K_2}\gamma_{12}\Delta_1\Delta_2.
\label{alpha}
\end{eqnarray}
We assume that $\gamma_{12}<0$.
The coefficient $\alpha$ has the dimension of the inverse of square of
the distance.
 
In the three-gap case, the Lagrangian for the phase part is given as
\begin{eqnarray}
\mathcal{L}&=& \frac{1}{4}\rho_1(\partial_t\theta_1)^2 
+ \frac{1}{4}\rho_2(\partial_t\theta_2)^2 
+ \frac{1}{4}\rho_3(\partial_t\theta_3)^2 \nonumber\\
&& -\frac{n_1}{8m_1}(\nabla\theta_1)^2 
 -\frac{n_2}{8m_2}(\nabla\theta_2)^2-\frac{n_3}{8m_3}(\nabla\theta_3)^2
\nonumber\\
&& -2\gamma_{12}\Delta_1\Delta_2\cos(\theta_1-\theta_2) 
 -2\gamma_{23}\Delta_2\Delta_3\cos(\theta_2-\theta_3) \nonumber\\ 
&& -2\gamma_{31}\Delta_3\Delta_1\cos(\theta_3-\theta_1),
\end{eqnarray}
where we assume that $\gamma_{ij}=\gamma_{ji}$. 
We define the total phase and phase
differences as follows.
\begin{eqnarray}
\phi &=& \theta_1+a\theta_2+b\theta_3, \\
\varphi_1 &=& \theta_1-\theta_2,\\
\varphi_2 &=& \theta_2-\theta_3 .
\end{eqnarray}
We choose $a$ and $b$ as $a=K_2/K_1$ and $b=K_3/K_1$.
We obtain the effective Lagrangian after integrating out the total
phase $\phi$ as
\begin{eqnarray}
\mathcal{L}_{\varphi} &=& 
\frac{\rho_1(\rho_2+\rho_3)}{4(\rho_1+\rho_2+\rho_3)}(\partial_t\varphi_1)^2
+ \frac{\rho_3(\rho_1+\rho_2)}{4(\rho_1+\rho_2+\rho_3)}(\partial_t\varphi_2)^2
\nonumber\\
&& +\frac{\rho_1\rho_3}{2(\rho_1+\rho_2+\rho_3)}\partial_t\varphi_1\partial_t\varphi_2
\nonumber\\
&& -\frac{K_1(K_2+K_3)}{4(K_1+K_2+K_3)}(\nabla\varphi_1)^2
-\frac{K_3(K_1+K_2)}{4(K_1+K_2+K_3)}(\nabla\varphi_2)^2
\nonumber\\
&& -\frac{K_1K_3}{2(K_1+K_2+K_3)}\nabla\varphi_1\cdot\nabla\varphi_2
\nonumber\\
&& -2\gamma_{12}\Delta_1\Delta_2\cos(\varphi_1)
-2\gamma_{23}\Delta_2\Delta_3\cos(\varphi_2) \nonumber\\
&& -2\gamma_{31}\Delta_3\Delta_1\cos(\varphi_1+\varphi_2). 
\end{eqnarray}

\subsection{Moduli approximation and mass of a kink and a monopole}

Let us consider the static kink in a two-gap superconductor and
examine the energy functional given by
\begin{equation}
E[\varphi] = A\int d^dx\left[ \frac{1}{2}(\nabla\varphi)^2-\alpha\cos(\varphi)\right].
\end{equation}
We consider a two-dimensional superconductor where there is a one-dimensional 
kink with variable $y$ being independent of the other space variable $x$.
The kink satisfies the equation
\begin{equation}
\frac{\partial^2\varphi}{\partial y^2}= \alpha\sin\varphi.
\end{equation}
A one-kink solution is
\begin{equation}
\varphi= \pi+2\sin^{-1}(\tanh(\sqrt{\alpha}(y-y_0))),
\label{onekink}
\end{equation}
with the boundary condition $\varphi\rightarrow 0$ as $y\rightarrow -\infty$
and $\varphi\rightarrow 2\pi$ as $y\rightarrow \infty$.
$y_0$ indicates the position where the kink exists and take any real value.
A set $\{y_0\}_{y_0\in{\bf R}}$ represents a moduli space of one-kink solution.
We here regard $y_0$ as a variable that depends on time $t$:
$y_0= y_0(t)$.  We obtain
\begin{equation}
\left( \frac{\partial \varphi}{\partial t}\right)^2 = 4\alpha
\frac{1}{\cosh^2(\sqrt{\alpha}(y-y_0))}\left( \frac{d y_0}{d t}
\right)^2.
\end{equation} 
Then the action for $y_0(t)$ is given by
\begin{equation}
S_{kink}[y_0] = \int dtdx A\frac{1}{2v^2}\left(\frac{\partial \varphi}{\partial t}\right)^2
=\int dt \frac{4A}{v^2}\sqrt{\alpha}\left(\frac{d y_0}{d t}\right)^2.
\end{equation}
This indicates that the kink motion exhibits a zero-energy mode (massless mode) with
the mass
\begin{equation}
M_{\alpha}= \frac{8A}{v^2}\sqrt{\alpha}L_x=\frac{4K_1K_2}{v^2(K_1+K_2)}
\sqrt{\alpha}L_x,
\end{equation}
where $L_x$ is the length of the kink in the $x$ direction.
The kink would move with a constant velocity $u$:
$y_0(t)= y_0(0)+ut$.
When two bands are equivalent in a simple case with $K_1=K_2=K=n/2m$
and $\rho_1=\rho_2=\rho_F$, $M_{\alpha}$ is given by
\begin{equation}
M_{\alpha}= 2\rho_F\sqrt{\alpha}L_x,
\end{equation}
with
\begin{equation}
\alpha= \frac{2d}{\pi^2}\frac{|\gamma_{12}|}{\rho_F}\frac{1}{\xi_0^2},
\end{equation}
where $d$ is the spatial dimension and $\xi_0$ is the coherence length
$\xi_0=\hbar v_F/(\pi\Delta)$.
$1/\sqrt{\alpha}$ indicates the spread of kink that is proportional
to $\xi_0$.
In two dimensions, $M_{\alpha}$ is proportional to the electron mass $m$,
\begin{equation}
M_{\alpha}= \frac{m}{\pi}\sqrt{\alpha}L_x.
\end{equation}
This massless mode is the sliding motion of kink\cite{tan17b}.

A fractional-flux vortex is regarded as a monopole in a superconductor
where the fractional vortex exists at the end of the kink, for instance, 
at $(x,y)=(x_0,0)$ and the other end is at the boundary of superconductor.
The phase of the second gap is represented as
\begin{equation}
\theta_2= \frac{1}{2}{\rm Im}\log(x-x_0+iy),
\end{equation}
and that of the first gap is $\theta_1=-\theta_2+\theta$ where $\theta$
is the rotation angle about the point $(x_0,0)$.
The gauge field corresponding to this phase configuration has a
singularity and represents a monopole in the phase space\cite{yan13}.
The phase difference is given by $\varphi=2\theta_1-\theta\equiv 0$ mod$2\pi$.
The fractional vortex with flux $\phi_0/2$ exists at $(x,y)=(x_0,0)$.
Let $x_0$ be dependent on time $t$: $x_0=x_0(t)$.  Then we have
\begin{equation}
\left(\frac{\partial\theta_1}{\partial t}\right)^2
= \frac{1}{4}\frac{y^2}{((x-x_0)^2+y^2)^2}
\left(\frac{d x_0}{dt}\right)^2.
\end{equation}
The action for $\varphi$ is given as
\begin{equation}
S_{mpole}[x_0]= \int dxdy\frac{A}{2v^2}\left(\frac{\partial\varphi}{\partial t}\right)^2
= \frac{1}{2}M_{mp}\left(\frac{dx_0}{dt}\right)^2,
\end{equation}
where the mass $M_{mp}$ is given by
\begin{equation}
M_{mp}= \frac{\pi}{2}\log\left(\frac{L_x}{a}\right)\cdot
\frac{\rho_1\rho_2}{\rho_1+\rho_2},
\end{equation}
where $a$ is a cutoff.
In two spatial dimensions, using $\rho_i=m_i/(2\pi)$, $M_{mp}$ is
proportional to the effective mass of electron:
\begin{equation}
M_{mp}= \frac{m^*}{4}\log\left(\frac{L_x}{a}\right),
\end{equation}
with $m^*=m_1m_2/(m_1+m_2)$.
A fractional-flux vortex, namely, the monopole exhibits a massless mode
with the mass proportional to $\log L_x$.

\subsection{Quantization of kink}

Let us examine the excitation modes of kinks in two-gap and three-gap
superconductors.
The kink solution has a continuous parameter called a moduli which
forms a space of solution.
Since the kink solution is a classical solution of the equation of motion,
there is a zero-energy mode in the tangent space of the classical 
solution.
We write the solution of kink as $\varphi= \varphi_0+\varphi^{(1)}$
where $\varphi_0$ is the kink state given in eq.(\ref{onekink}) and
$\varphi^{(1)}$ represents a fluctuation to the classical kink solution.
$\varphi^{(1)}$ describes excitation modes of kink in a two-gap superconductor.
We obtain the energy functional $E[\varphi]$ up to the second order
of $\varphi^{(1)}$.
$\varphi^{(1)}(x)$ satisfies the eigenvalue equation given as
(we use $x$ as a variable)
\begin{equation}
\left( -\frac{\partial^2}{\partial x^2}+\alpha\cos\varphi_0\right)
\varphi^{(1)}= \lambda\varphi^{(1)}.
\end{equation}
Since the eigenvalue is zero or positive, we put $\lambda=\omega^2$
where $\omega$ (times $A$) indicates the excitation energy.
The eigenfunctions and eigenvalues are given by
\begin{eqnarray}
\varphi^{(1)}_0&=& \frac{1}{\cosh(\sqrt{\alpha}x)}, ~~~~ \lambda_0=0,\\
\varphi^{(1)}_1&=& \tanh(\sqrt{\alpha}x), ~~~~ \lambda_1=\alpha, \\
\varphi^{(1)}_q&=& e^{iqx\sqrt{\alpha}}(q+i\tanh(\sqrt{\alpha}x)),
~~~ \lambda_q=(q^2+1)\alpha,
\end{eqnarray}
where we set $x_0=0$ for simplicity and $q$ in the third equation is
quantized according to the boundary condition.
We adopt that the system is the box of length $L$ and we impose the
periodic boundary condition at $x=-L/2$ and $L/2$:
\begin{equation}
q_nL\sqrt{\alpha}+2\eta(q_n)=2n\pi,
\end{equation}
where $\eta(q)$ is the phase difference when $x$ is large given by
$\tan\eta(q)=1/q$.
The parameter $q$ is quantized as $q=q_n$ for $n=0,1,\cdots$.
$\varphi^{(1)}_0$ represents the zero energy mode, and $\varphi^{(1)}_1$
and $\varphi^{(1)}_q$ are quantized excitation modes.

Let us consider excitation modes of kink in the three-gap case. 
We examine the case where two bands 1 and 3 are equivalent and we set
$\gamma_{12}=\gamma_{23}$\cite{yan13}, $\Delta_1=\Delta_3$ and
$K_1=K_3$.  In this case we have $\varphi_1=\varphi_2\equiv \varphi$.
The energy functional is written as
\begin{equation}
E[\varphi]= \frac{1}{g}\int d^dx\left[ \frac{1}{2}(\nabla\varphi)^2
-\alpha\left( \cos\varphi+\frac{u}{2}\cos(2\varphi)\right)\right],
\end{equation}
where $u=\gamma_{13}\Delta_1/(\gamma_{12}\Delta_2)$ and $g=A^{-1}$
given as
\begin{equation}
\frac{1}{g}= \frac{K_1(K_1+2K_2)}{2K_1+K_2}.
\end{equation}
The parameter $\alpha$ is 
\begin{equation}
\alpha= -4\frac{2K_1+K_2}{K_1(K_1+2K_2)}\gamma_{12}\Delta_1\Delta_2.
\end{equation}
We assume that $\gamma_{12}<0$ and $u>-1/2$.
We consider a one-dimensional kink with variable $x$ as before and the 
solution is given by
\begin{equation}
\varphi_{kink}=\varphi_0= \cos^{-1}\left(
\frac{2\sinh^2(s(x-x_0))}{\cosh^2(s(x-x_0))+2u}
-1\right),
\end{equation}
where we put $s=\sqrt{\alpha(1+2u)}$.
When we regard the parameter $x_0$ as a time-dependent parameter,
$x_0(t)$ represents a gapless mode with the mass
\begin{equation}
M_{\alpha}= 2\rho_F^*\sqrt{\alpha(1+2u)}
\left( 1+\frac{1}{\sqrt{2u(1+2u)}}\cosh^{-1}(\sqrt{1+2u})\right)L_{k},
\end{equation}
where $\rho_F^*$ is the effective density of states and $L_k$ is
the length of the kink.

The kink solution is written as $\varphi=\varphi_0+\varphi^{(1)}$ as
before where $\varphi^{(1)}$ represents a correction to the kink
state.  We examine the following eigenvalue equation:
\begin{equation}
\left( -\frac{\partial^2}{\partial x^2}+\alpha(\cos\varphi+2u\cos(2\varphi))\right)
\varphi^{(1)}(x)= \lambda\varphi^{(1)}(x).
\end{equation}
We change the variable $x$ to $z=s(x-x_0)$ to the eigenvalue equation
as follows:
\begin{equation}
\left( -\frac{d^2}{dz^2}+1-\frac{2+16u}{\cosh^2 z+2u}
+\frac{16u(1+2u)}{(\cosh^2 z+2u)^2}\right) \varphi^{(1)}(z)
= \nu\varphi^{(1)}(z),
\label{ueigen}
\end{equation}
where $\nu=\lambda/s^2$.
The zero-energy mode is easily found as
\begin{equation}
\varphi^{(1)}_0(x)= \frac{\cosh(s(x-x_0))}{\cosh^2(s(x-x_0))+2u},
\end{equation}
with the eigenvalue $\nu_0=\lambda_0=0$.
It is not easy to obtain an exact expression for excited states
with positive eigenvalues.
An approximate solution of the first excited state for small $u$
and large $x$ is obtained as
\begin{equation}
\varphi^{(1)}_1(z) \simeq \frac{\sinh(z)}{\cosh(z)}
-4u\frac{\sinh(z)}{\cosh^3(z)}.
\end{equation}
The eigenvalue is $\nu_1\simeq 1$, namely, $\lambda_1\simeq s^2$. 
The third and higher excited states have continuous spectra when
the system size $L$ is very large.  Under the same condition,
$\varphi^{(1)}_q$ will be approximated as
\begin{equation}
\varphi^{(1)}_q(z)\simeq e^{iqsx}(q+i\varphi^{(1)}_1(z)),
\end{equation}
with the eigenvalue $\nu_q\simeq q^2+1$.
 
We solve the eigen-equation in eq.(\ref{ueigen}) numerically by
reducing it to a tridiagonal matrix on a lattice. 
For $u=0$ we have $\nu=0, 1$ and $q_n^2+1$ $(n=0,1,\cdots)$, and
for $u>0$, $\nu=0, \nu_1$ and $q_n(u)^2+1$ $(n=0,1,\cdots)$.
We found $\nu_1\leq 1$ for $u\geq 0$ and $q_n(u)$ is almost
independent of $u$.  
We here mention that $\lambda_1/\alpha=\nu_1s^2/\alpha=\nu_1(1+2u)\geq 1$
although $\nu_1\leq 1$.
We show the zero energy mode as a function of $x$ in Fig.4
where $\xi_{kink}=1/s$ is the characteristic length of kink and
is the unit of length. 
This mode is localized near $x=x_0$ and its spreading width is
about $\xi_{kink}$.
The excitation mode is a spread state. The first and second excited
states are shown in Fig.5 and Fig.6, respectively.
We show eigenvalues $\nu_n=\lambda_n/s^2$ as a function of $u$ in Fig.7.
The first excited state shows strong dependence on $u$, and other
states, however, are mostly independent of $u$.
The size dependence of eigenvalues is shown as a function of $1/L$ in Fig.8.
The eigenvalues of higher excited modes constitute a continuous spectrum
when $L$ is large.

\begin{figure}[htbp]
\begin{center}
\includegraphics[height=7.0cm]{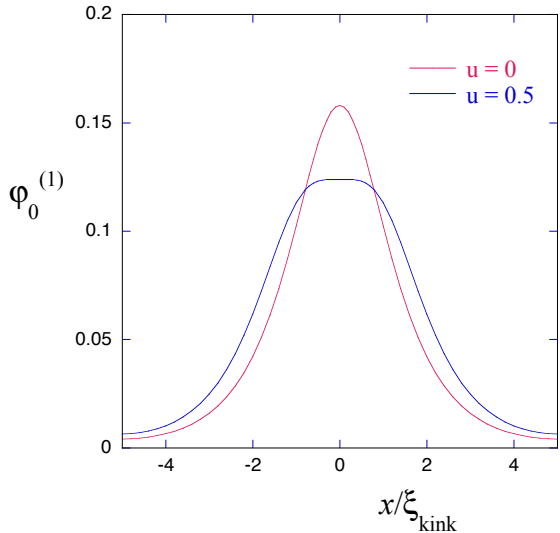}
\caption{
Zero-energy mode as a function of $x$ with $x_0=0$ for $u=0$ and $0.5$.
}
\label{kinkmode0}
\end{center}
\end{figure}

\begin{figure}[htbp]
\begin{center}
\includegraphics[height=7.0cm]{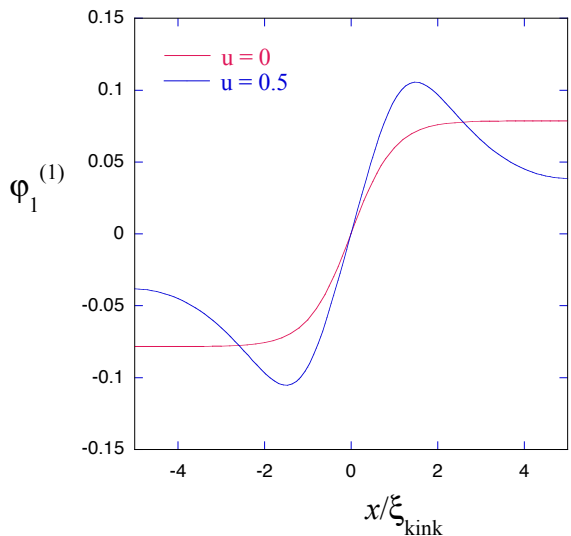}
\caption{
First excited mode as a function of $x$ with $x_0=0$ for $u=0$ and $0.5$.
}
\label{kinkmode1}
\end{center}
\end{figure}

\begin{figure}[htbp]
\begin{center}
\includegraphics[height=7.0cm]{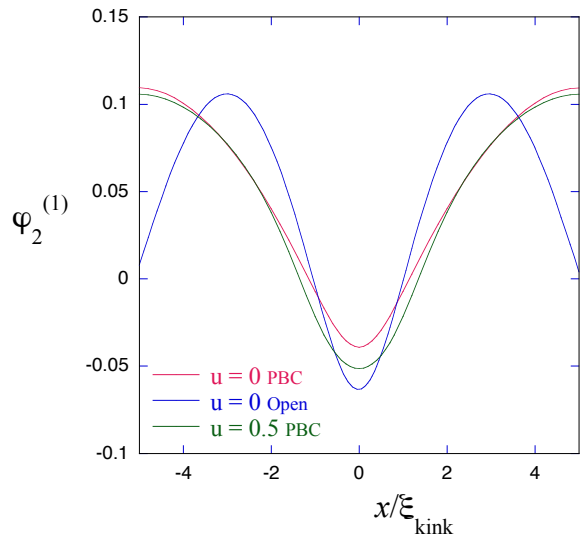}
\caption{
Second excited mode as a function of $x$ with $x_0=0$ for $u=0$ and $0.5$.
PBC and Open indicate the periodic and open boundary condition,
respectively.
The second and higher excited states depend on the boundary condition.
}
\label{kinkmode2}
\end{center}
\end{figure}

\begin{figure}[htbp]
\begin{center}
\includegraphics[height=6.5cm]{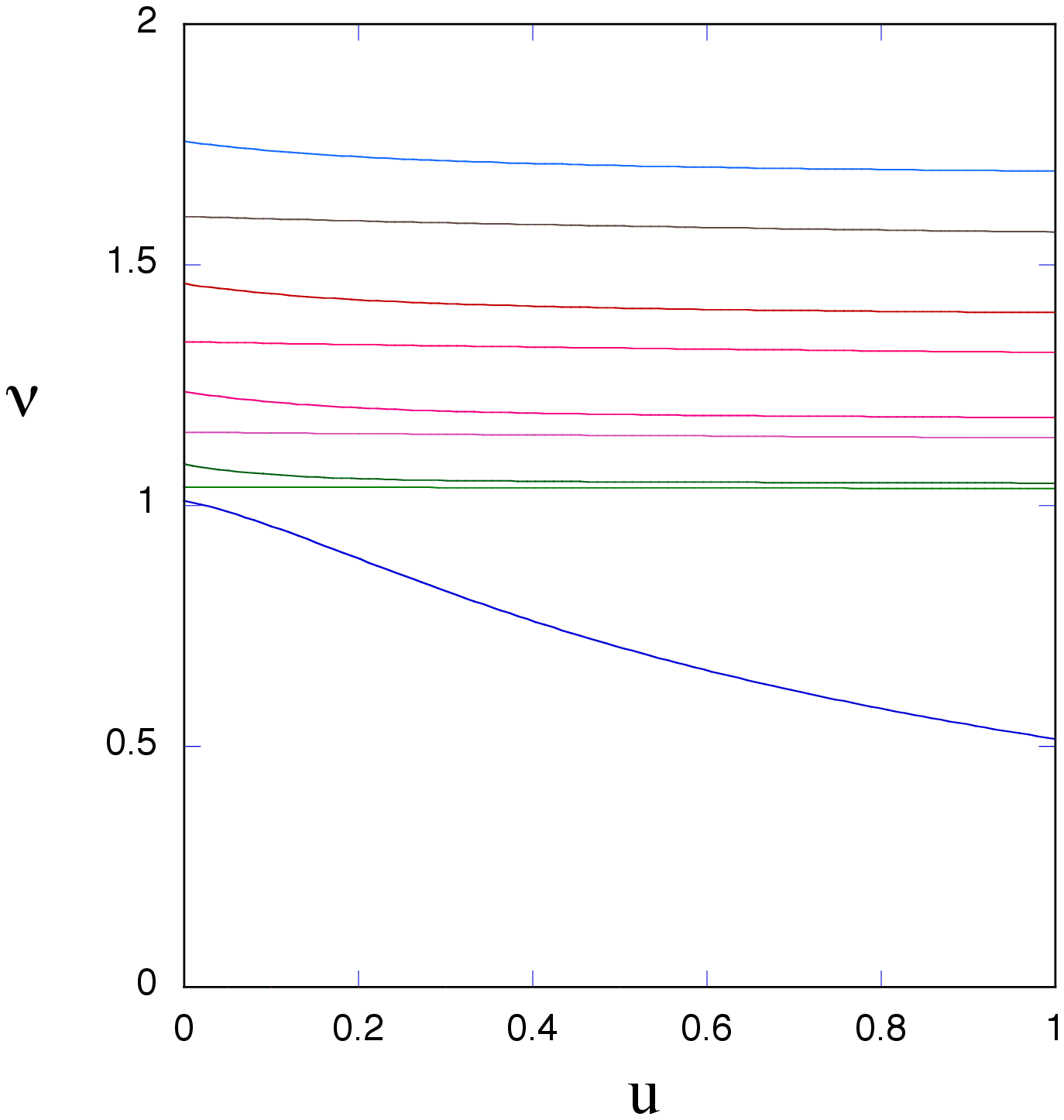}
\caption{
Eigenvalues $\nu_n=\lambda_n/s^2$ as a function of $u$ 
($n=0,1,\cdots,9$ from the bottom) where $\nu_0=0$.
}
\label{kinku}
\end{center}
\end{figure}

\begin{figure}[htbp]
\begin{center}
\includegraphics[height=6.5cm]{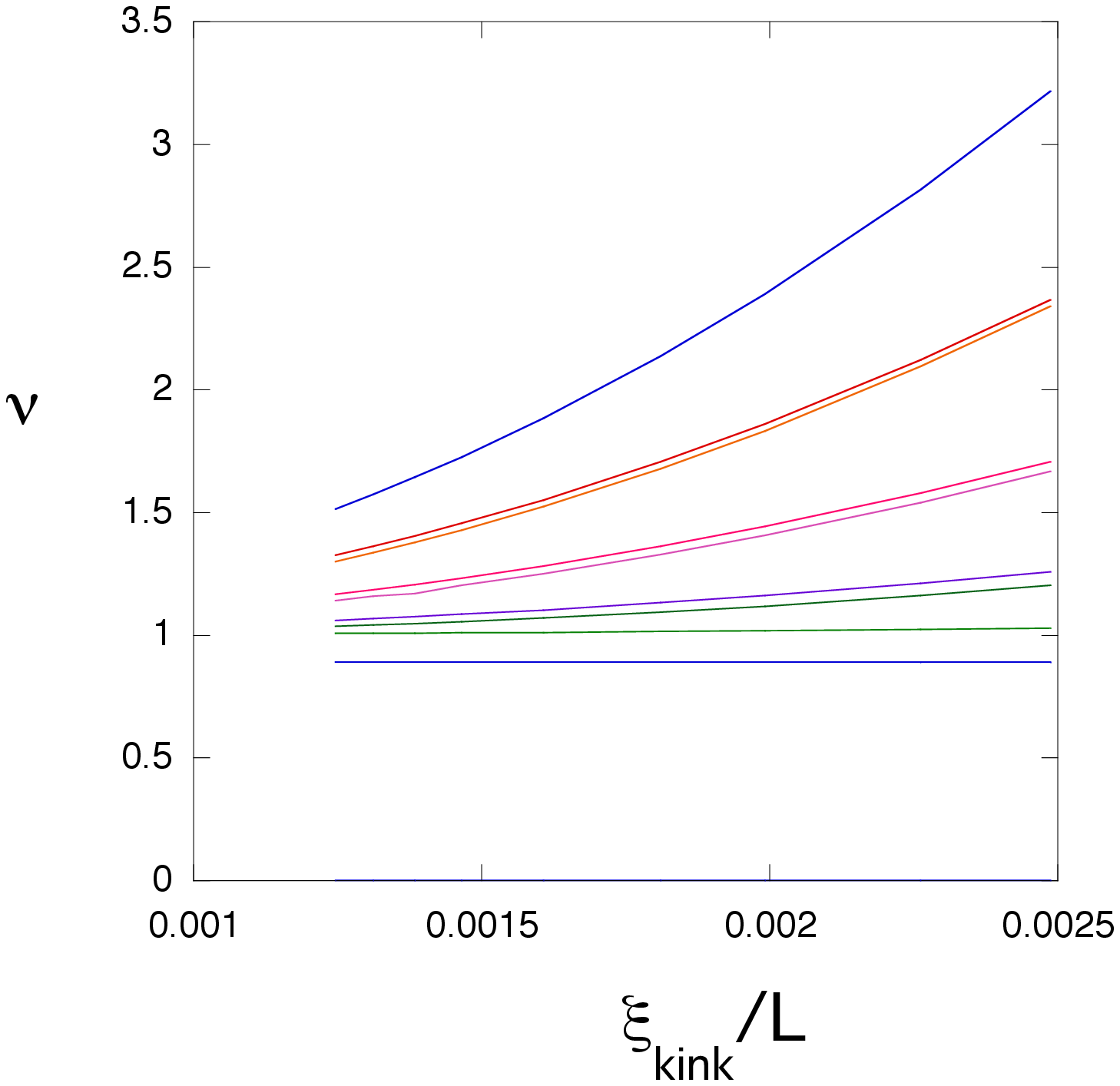}
\caption{
Eigenvalues $\nu_n=\lambda_n/s^2$ as a function of $1/L$ 
($n=0,1,\cdots,9$ from the bottom) where $\nu_0=0$.
}
\label{kinkn}
\end{center}
\end{figure}

\subsection{Energy of Kink}

Here let us discuss the energy of kink.
The energy of the solution $\varphi_0$ is given by
\begin{equation}
E[\varphi_0]= \frac{1}{g}\int dx \Bigg[ 
\frac{1}{2}\left(\frac{d\varphi_0}{dx}\right)^2
-\alpha\left( \cos\varphi_0+\frac{u}{2}\cos(2\varphi_0)
-1-\frac{u}{2}\right)\Bigg],
\end{equation}
where a constant $-(1+u/2)$ is introduced to subtract the divergence.
This is evaluated as
\begin{equation}
E[\varphi_0] = \frac{1}{g}4s\left( 1+\frac{1}{\sqrt{2u(1+2u)}}
\cosh^{-1}(\sqrt{1+2u})\right).
\end{equation}

The excitation modes are approximated as a set of harmonic oscillators.
The field $\varphi$ is written as
\begin{equation}
\varphi= \varphi_0(x-x_0(t))+\sum_nc_n(t)\varphi^{(1)}_n(x-x_0(t)),
\end{equation}
where we adopt that $x_0$ and the coefficients $\{c_n\}$ are time dependent.
$x_0(t)$ and $c_0(t)$ give the zero-energy mode.
The Lagrangian for $c_n$ with $n\neq 0$ is given by
\begin{equation}
L[c_n]= \frac{1}{2gv^2}\sum_n\left( \dot{c}_n^2-v^2\omega_n^2c_n^2\right),
\end{equation}
where we set $\lambda_n=\omega_n^2$.  The excitation modes are described
by harmonic oscillators with frequencies $v\omega_n$. 
The corrections to the energy from quantized modes are written as
\begin{equation}
E_{kink}= E[\varphi_0]+\left(N_1+\frac{1}{2}\right)v\omega_1
+\sum_{q_n}\left(N_{q_n}+\frac{1}{2}\right)v\omega_{q_n},
\end{equation}
where $N_1$ and $N_{q_n}$ are integers, and
$v^2=(\rho_1+\rho_2)/(\rho_1\rho_2)\cdot K_1K_2/(K_1+K_2)$ in
the two-gap case ($u=0$) and
\begin{equation}
v^2=\frac{2\rho_1+\rho_2}{\rho_1(\rho_1+2\rho_2)}
\frac{K_1(K_1+2K_2)}{2K_1+K_2},
\end{equation} 
in the three-gap case.  We put
$\omega_1=\sqrt{\lambda_1}=s\sqrt{\nu_1}$ and
$\omega_{q_n}=\sqrt{\lambda_{q_n}}=s\sqrt{\nu_{q_n}}$.
According to numerical calculations in Fig.7, $\nu_{q_n}$ is mostly
independent of $u$.  Thus, we use the formula for $u=0$:
 $\nu_{q_n}\simeq q_n^2+1$ for $n=0,1,2,\cdots$.
$q_n$ is determined from the boundary condition given by
\begin{equation}
q_nLs+2\eta(q_n)=2n\pi\equiv k_nL.
\end{equation}
Then, $E_{kink}$ is approximated as
\begin{equation}
E_{kink}\simeq E[\varphi_0]+\left(N_1+\frac{1}{2}\right)vs\sqrt{\nu_1}
 +\sum_{q_n}\left( N_{q_n}+\frac{1}{2} \right)vs\sqrt{q_n^2+1}.
\end{equation}
The ground state energy of kink is given as
\begin{equation}
E_{kink,0}\simeq E[\varphi_0]+\frac{1}{2}vs\sqrt{\nu_1}
+\frac{1}{2}\sum_{q_n}vs\sqrt{q_n^2+1}.
\end{equation}
Since the summation with respect to $q_n$ diverges, we subtract 
$\sqrt{k_n^2+s^2}$ from $s\sqrt{q_n^2+1}$.  We obtain
for large $L$
\begin{eqnarray}
 && \frac{1}{2}v\sum_{k_n}\left( \sqrt{(q_ns)^2+s^2}-\sqrt{k_n^2+s^2}\right) \nonumber\\
= && -\frac{L}{4\pi}v\frac{2}{L}\int_{-\infty}^{\infty}dk
\frac{k}{\sqrt{k^2+s^2}}\eta(k) \nonumber\\
= && -\frac{vs}{\pi}-\frac{vs}{2\pi}\int_{-\infty}^{\infty} dk
\frac{1}{\sqrt{k^2+s^2}},
\label{diverg}
\end{eqnarray}
where we used partial integration and $\eta(k)\sim s/k$ for large $|k|$.
The last integral diverges, which is removed by the renormalization
of $\alpha\propto s^2$.
In fact, the same divergence appears from the interaction term
$-(\alpha/g)\cos\varphi$. 
The one-loop contribution is given by\cite{yan16,yan17,yan18}
$(\alpha/g)(\langle\varphi^2\rangle/2)\cos\varphi$.
This is estimated by using
\begin{equation}
\langle\varphi^2\rangle= \int \frac{d\omega dk}{(2\pi)^2}
\frac{g}{\omega^2/v^2+k^2+s^2}= 
\frac{gv}{4\pi}\int_{-\infty}^{\infty}dk\frac{1}{\sqrt{k^2+s^2}}.
\end{equation}
$\cos\varphi$ is approximated as 
\begin{equation}
\langle\cos\varphi_0-1\rangle = -2\int_{-\infty}^{\infty}dx
\frac{1}{\cosh^2(sx)}= -\frac{4}{s},
\end{equation}
by subtracting 1 form $\cos\varphi_0$ to regularize the integral.
This gives the same correction as in eq.(\ref{diverg}) and is cancelled by 
introducing the counter term.
For the interaction $-(\alpha/g)[\cos\varphi+(u/2)\cos(2\varphi)]$,
the contribution to the ground-state energy is given by evaluating
\begin{eqnarray}
 && \langle \cos(\varphi_0)+2u\cos(2\varphi_0)-1-2u\rangle \nonumber\\
= && (1+2u)\int_{-\infty}^{\infty}dx\Bigg[ -\frac{2+16u}{\cosh^2(sx)+2u}
+\frac{16u(1+2u)}{(\cosh^2(sx)+2u)^2}\Bigg]\nonumber\\
= && \frac{2}{s}(1+2u)\Bigg[ -4+\frac{2}{\sqrt{2u(1+2u)}}\cosh^{-1}\sqrt{1+2u}
\Bigg]\nonumber\\
= && -\frac{4}{s}(1+2u)\left( 1+O(u)\right).
\end{eqnarray}
Then the ground-state energy of kink is
\begin{equation}
E_{kink,0}\simeq E[\varphi_0]+\left( \frac{1}{2}\sqrt{\nu_1}
-\frac{1}{\pi}\right)vs.
\end{equation}

\section{Summary}

The observation of fractional-flux vortices in a layered
superconductor indicates that there are indeed kinks in the phase
space of a multi-component superconductor\cite{tan18}.
Thus, it is important to investigate physical properties of kinks
in superconductors.

We investigated quantized excitation modes of the kink in a two-gap and
a three-gap superconductor.  There are zero-energy mode (massless mode) and
quantized excitation modes with energy gaps proportional to $s\propto \sqrt{\alpha}$.
Second and higher excitation modes have almost continuous spectra when
the system size $L$ is large.
Since $\alpha$ is proportional to the Josephson coupling, properties
of excitation modes are determined by the Josephson coupling
between different gap states.
Although the energy of kink has a divergence, it is regularized
by the renormalization of coupling constants in the Lagrangian.
The zero-energy mode has a peak at the position of the kink,
which indicates a fluctuation of the amplitude of domain wall at
the position of the kink.
We expect that this mode can be observed by some experimental
equipments such as scanning tunneling spectroscopy.

There are two massless modes so far; one is the massless mode of the kink
and the other is that of sliding motion of the kink.
The former is a small oscillation mode of the kink, which may be called
the ripple mode.
A fractional vortex itself also exhibits a massless mode as a motion
of a monopole.

We also examined a stability of fractional-flux vortices in a
multiband superconductor.
In a two-band superconductor, a fractional-flux vortex is unstable
because of the energy cost of kink.
Two fractional vortices connected by a kink are glued to be a
vortex with the unit flux.  
In a real superconductor, however, fractional-flux vortices may be
stabilized due to some vortex pinning effect and magnetic interaction
between vortices and external magnetic field outside the superconductor. 
An interesting possibility of stable fractional vortices was found in a
three-gap superconductor with time-reversal symmetry
breaking.  It was shown that fractional vortices can exist
on the transverse kink that connects two states with the same energy.
The unit flux is decomposed into two vortices with flux $\phi_0/3$
and $2\phi_0/3$ along the transverse kink.
This is a vortex pinning by kinks.
The vortex pinning by kinks indicates a possibility of high
critical current of a superconducting wire.

We thank T. Nishio, S. Arisawa and H. Yamamori for useful discussions.
This work was supported by Grant-in-Aid from the
Ministry of Education, Culture, Sports, Science and
Technology of Japan (Grant No. 17K05559).



\end{document}